\documentclass[namedreferences,hyperref,optionalrh]{spr-sola}
\usepackage{graphicx}        
\usepackage{color}           

\usepackage{url}
\usepackage{xcolor}
\usepackage{hyperref}
\hypersetup{
    colorlinks,
    linkcolor={red!50!red},
    citecolor={red!50!red},
    urlcolor={red!80!red}
}

\chardef\us=`\_

\begin{document}

\begin{frontmatter}
\title{LONG-LIVED SUNSPOTS IN HISTORICAL RECORDS: 
A CASE STUDY ANALYSIS FROM 1660 TO 1676}

\author[addressref=aff1,corref,email={ned@geo.phys.spbu.ru}]{\inits{N.}\fnm{Nadezhda}~\snm{Zolotova}\orcid{0000-0002-0019-2415}
}\author[addressref={aff2},email={mikhail.vokhmianin@oulu.fi}]{\inits{M.V.}\fnm{Mikhail}~\lnm{Vokhmyanin}\orcid{https://orcid.org/0000-0002-4017-6233}}

\address[id=aff1]{St. Petersburg State University, Universitetskaya nab. 7/9, 198504 St. Petersburg, Russia}

\address[id=aff2]{Space Climate Group, Space Physics and Astronomy Research Unit, University of Oulu, Oulu, Finland}

\runningauthor{  }
\runningtitle{  }

\begin{abstract}

Sunspot engravings and measurements in 1660\,--\,1676 are analyzed to retrieve sunspot area and heliocoordinates. Based on these data, we revise the \citeauthor{1997rscc.book.....H} (The role of the sun in climate change, 
\citeyear{1997rscc.book.....H}) hypothesis of long-lived sunspots during the Maunder minimum as a sign of weakened convection. Historical reports also clarify what each observer defined as a sunspot and the purpose of the observations. The reconstructed longitudes of sunspots allow us to evaluate the rotation rate, revealing that the historical rotation profile resembles that of long-lived sunspot groups in the modern era.

\end{abstract}
\end{frontmatter}

\section{Introduction}
     \label{S-Introduction} 

One of the earliest references to suppressed solar activity in the seventeenth century comes from \citet{1710RSPT...27..270D} and \citet{Hausen_1726}, who noted that from 1660 to 1670/1671 and from 1676 to 1684, the Sun was nearly devoid of sunspots. Nowadays, the period from 1645\,--\,1715 is known as a representative grand solar minimum, attracting significant attention to historical solar studies \citep{2020LRSP...17....1A, 2024SoPh..299...45B, 2021ApJ...922...58C, 2021SoPh..296...59C, 2022ApJ...933...26C, 2024ApJ...968...65C, 2024JSWSC..14....9C, 2022SoPh..297...79I, 2023MNRAS.523.1809I, 2018smhr.book.....G,  2020ApJ...890...98H, 2021ApJ...919....1H, 2024MNRAS.528.6280H, 2021NatSR..11.5482M, 2024AN....34530131N}.

\citet{Spoerer1889} compiled a list of sunspots and their latitudes for the period 1672\,--\,1713. \citet{1997rscc.book.....H} identified those that were long-lived between 1672 and 1700. They also hypothesized that a sunspot observed between 9 May and 7 August 1660\footnote{There is a typo in \citet{1997rscc.book.....H}: 1661 instead of 1660.} might have been the same sunspot throughout this period. Based on the dates of sunspot reports, they suggested that this sunspot group could have persisted even longer, potentially from late February to early August 1660 spanning seven solar rotations. 

\citet{1997rscc.book.....H} estimated that approximately 10\% of the sunspots (2 out of 23) observed between 1672 and 1700 lasted for at least four rotations. If sunspots during this period were indeed longer-lived than the present-day sunspots, it could suggest weaker convection in the late 1600s, as proposed by \citep{1976IAUS...71....3P}.

In this study, we analyze historical reports of long-lasting active regions. Their heliocoordinates have been reconstructed and are provided in the Electronic Supplementary Materials. Long-lived sunspots enable more accurate assessments of the sunspot rotation rate, though this requires precise knowledge of their longitudinal positions. The method we use to retrieve the orientation of the heliographic grid is described in \citet{2018SoPh..293...31V}. To define the solar ephemeris, we employ the French planetary theory VSOP87 \citep{1988A&A...202..309B, 1991aalg.book.....M}. Longitudes are measured from the zero meridian at Greenwich Noon on 1 January 1854 and rotate with a sidereal period of 25.38 days, as conventionally fixed by Carrington.

\section{Year 1660}
\label{S-1660}

\citet{1671RSPT....6.2216C} cites notes by Boyl, presumably Robert Boyle, best known for Boyle's law: ``Friday, April 27, 1660 [7 May in the Gregorian calendar], about 8 of the clock in the Morning, there appear's Spot in the lower limb of the Sun a little towards the South of its AEquator, which was entred about 1/40 of the Diameter of the Sun, it itself being about 1/165 [the last digit is poorly printed, but we interpret it as 5] in its shortest Diameter, of that of the Sun; its longest, about 1/40 of the same. It disappear'd upon Wednesday Morning (May 9th) [19 May Greg.] though we saw it about day before about 10 in the morning to be near about the same distance from the Westward limb a little South of its aequator, that it first appear'd to be from the Eastward limb, a little South also of its aequator, It seem'd to move faster in the middle of the Sun then towards the limb. It was a very dark spot almost of a quadrangular form, and was enclosed round with a kind of duskish cloud, much in this form and in this proportion to the Spot [the text is supplemented by a schematic sketch of the sunspot]. We first observ'd this very same Spot both for figure, color and bulk, to be re-enter'd the Sun May 25th [4 June Greg.], when it seem'd to be in a part of the same line it had formerly traced; and was enter'd about 4/33 of its Diameter about 7 of the clock in the afternoon. At the same time there appear'd another Spot, which was just entred and appear'd to be entred not above 1/132 part of the Sun's diameter. It appear'd to be longest towards the North and South, and shortest towards the East and West. There seem'd to be dispers'd about it divers small clouds here and there.". These observations were reported to have been conducted using an excellent telescope.

\begin{figure}    
\centerline{\includegraphics[width=1\textwidth,clip=]{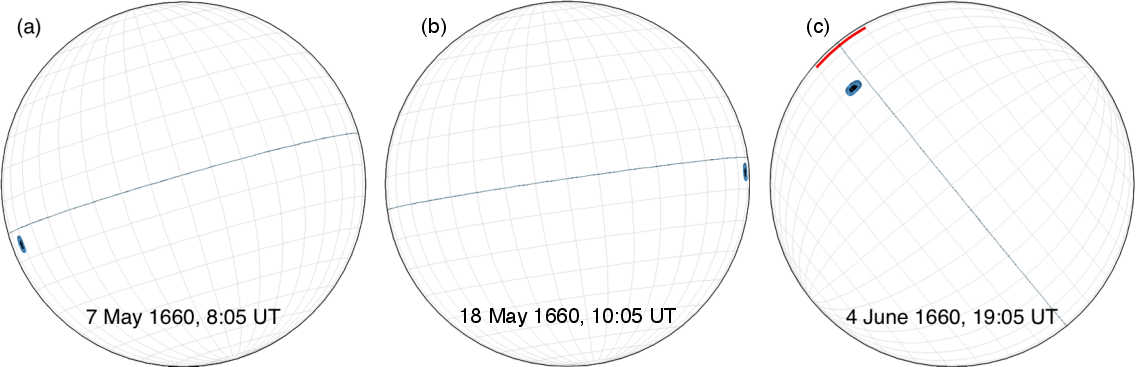}}
\small
        \caption{Plausible position of the two sunspots based on Boyle's note. \textit{Red arc} shows the possible latitude-longitude range where the second sunspot could be located.}
\label{Fig1}
\end{figure}

Using the sketch and measurements provided, we mapped the sunspots onto the solar disk (Figure~\ref{Fig1}). For the first sunspot (indicated in blue), we arbitrary interpreted the description ``a little towards the South of its equator" as $-5^{\circ}$ in latitude. Since no specific latitudinal constraint was provided for the second active region, we initially assumed its latitude to fall within $\pm30^{\circ}$. However, considering the reports by Johannes \citet{Hevelius_1679}, we narrowed the latitude range to $\pm10^{\circ}$ (Figure~\ref{Fig1}c, red arc). The Electronic Supplementary Materials include the reconstructed sunspot parameters derived from these interpretations (Figure~\ref{Fig1}). Note that the observations by Boyle and Hevelius share a common group numbering (G). Also note that Boyle referred to the sunspot's umbra as the ``spot", while describing its penumbra as a ``duskish cloud".

Precise measurements of the first sunspot's position on 7 May and 4 June 1660 yield a sidereal sunspot rotation rate of $14.4 \pm 0.1^{\circ}$ d$^{-1}$. This implies that on 18 May 1660 the sunspot was located at approximately $32^{\circ}$ in longitude (Figure~\ref{Fig1}b). 

Speculating that the sunspots or sunspot groups could have persisted for seven solar rotations due to weakened convection, we modeled the transit of two objects across the solar disk between February and August 1660. The rotation rate for the first sunspot as previously mentioned, is $14.4 \pm 0.1^{\circ}$ d$^{-1}$. For the second sunspot, whose rotation rate is unknown, we applied the rotation law derived by \citet{1986A&A...155...87B}: $\omega(B)=\omega_0+2.87\sin^2B$, where $B$ represents the heliographic latitude, and $\omega_0$ is set to $14.4 \pm 0.1^{\circ}$ d$^{-1}$.

\begin{figure}    
\centerline{\includegraphics[width=1\textwidth,clip=]{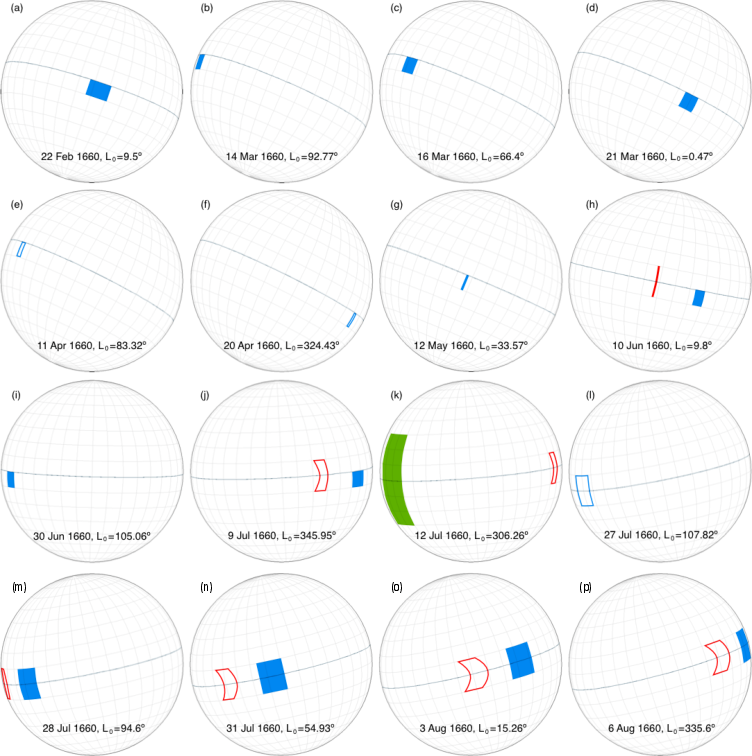}}
\small
        \caption{Estimated range of sunspot positions from 22 February to 6 August 1660. The \textit{blue}, \textit{red}, and \textit{green} regions represent the three sunspot groups. Areas without fill indicate positions that were not confirmed by the observations. $L_0$ denotes the heliographic longitude of the apparent disk center as seen from Gdansk, Poland.
        }
\label{Fig2}
\end{figure}

The colored regions in Figure~\ref{Fig2} represent the estimated latitude-longitude ranges, where the centers of the sunspot groups could have been located. The reference points for these estimations are Boyle's observations on 7 May and 4 June 1660 for the first (in blue) and the second (in red). To match the assumed range of sunspot positions with observations made by \citet{Hevelius_1679} in 1660, Figure~\ref{Fig2} shows the heliographic grid as it was seen in Gdansk, Poland, at local Noon (10:45 UT). However, it is not strictly that the sunspots were observed at Noon. 

In his book \textit{Machina Coelestis}, Hevelius provided astrometric and solar observations in tabular form. The solar observations primarily consist of midday altitude measurements, with less focus on sunspot reports and solar diameter measurements. The tables also document the equipment used, weather conditions, and the quality of the observations. Notably, only an azimuthal quadrant or horizontal quadrant is mentioned in these records. If a telescope was not utilized, the observer would have been limited to detecting large sunspots. However, Hevelius made a single mention of observing faculae, which suggests that a refracting telescope was employed.

Further, we rely on the translation of the original Latin text by \citet{2015SoPh..290.2719C}. On 22 February 1660, Hevelius reported a notable round spot that appeared in the middle of the solar disk, with a smaller spot that corresponds to the assumed range of sunspot positions shown in Figure~\ref{Fig2}a. The sunspot would have been exactly at the center, given a rotation rate of $14.45^{\circ}$ d$^{-1}$. This faster rotation, compared to the $14.4^{\circ}$ d$^{-1}$ rate, could be attributed to uncertainties or to the tendency of the rotation rate of long-lived recurrent sunspot groups to slow down over time \citep[Figure~9 therein]{1998A&A...332..748P}. In contrast, \citet{2023MNRAS.519.5315K} inferred that the rotation rate remains constant throughout the evolution of an active region. Furthermore, the area-weighted center of an active region may exhibit its own motion as the region evolves, introducing additional uncertainties in rotation rate \citep{1993ASPC...46..123P}.

On 26 February 1660, Hevelius reported that the larger spot had shrank and the smaller one had vanished. On 29 February, faint faculae and umbrae were observed\footnote{Notice that \citet{Hevelius_1647} used the term ``umbra" for many phenomena, e.g. to define dark regions between sunspots or in facular regions \citep{2019ApJ...886...18C}.}. According to our calculations, the large sunspot should have already been a few to $15^{\circ}$ behind the western limb on that day. Therefore, Hevelius could have observed smaller following spot(s), if any were behind the preceding spot, or it could have been another active region.

On 16 March 1660, Hevelius observed the spot along with two smaller ones near the eastern limb. He also assumed that the sunspots showed up on 13 or 14 March. This report is consistent with our calculations (Figures~\ref{Fig2}b and c). On 13 March, the sunspot, rotating at $14.3^{\circ}$ d$^{-1}$, is estimated to have been at the very edge of the solar disk. If it had been rotating faster, the spot would have still remained behind the limb. If the rotation rate exceeded $14.5^{\circ}$ d$^{-1}$, the sunspot would still have been behind the limb even on 14 March. On 17 March, Hevelius reported that the major sunspot had grown and was followed by five smaller and faint spots. On 21 March 1660, the spots were observed in the western quadrant. Finally, on 28 March, no spots were reported. Both entries align with our proposed sunspot positions: Figure~\ref{Fig2}d and on 28 March, the sunspots were $19-27^{\circ}$ beyond the limb.

On 11 and 20 April 1660, Hevelius described the solar disk as being blank. According to our calculations, the supposedly long-living sunspot was located near the eastern and western limbs, respectively (Figures~\ref{Fig2}e and f; here and below, in cases where the assumed range of sunspot positions is not confirmed by observations, we have omitted the color fill). Since the observations on 11 April were made diligently, as Hevelius noted in the relevant column of his table, we can speculate that the sunspot group lived only two solar rotations (February and March 1660). Moreover, at the beginning of the second rotation, Hevelius reported that the spot had enlarged, which is unlikely for a spot that has persisted for more than one rotation. This could be interpreted as the emergence of a new portion of magnetic flux.

On 12 May 1660, Hevelius stated a conspicuous round spot near the center of the Sun, which had already passed the disk center by the next day. He also suggested that it crossed the eastern limb around 6 May. By 15 May, the spot had shrunk, and by 18 May, it was close to the western limb. The following day, the spot had already exited the solar disk. These reports align precisely with Boyle's observations (Figures~\ref{Fig1}a and b) and our estimations: on 6 May, the sunspot was a few degrees from the eastern limb; on 12 May, it was near the disk center (Figure~\ref{Fig2}g); and by 19 May, it had just passed the limb.

On 2 June 1660, Hevelius reported that no sunspots were detected. However, according to our assumptions, the center of the active region should have been located at the very edge of the eastern limb.

On 10 June 1660, Hevelius described notable sunspots: three near the center of the Sun and one in the western quadrant. This observation agrees with Boyle's observations and our calculations: the center of the first sunspot was in the western quadrant, while the second active region was near the disc center (Figure~\ref{Fig2}h, blue and red, correspondingly). Since the exact position of the second spot relative to the Equator is unkown, the red stripe occupies both hemispheres. Hevelius also reported sunspots on 11 and 12 June, and by 16 June, no spots were observed. We estimate that the center of the second active region would have been at the very edge of the solar disk on 16 June.

On 9 July 1660, Hevelius noticed a large round sunspot, along with a smaller one near the southwestern limb. He suggested the group had entered the solar disk about 10 days earlier. This description matches the range of sunspot positions that we have marked in blue (Figures~\ref{Fig2}i and j). On 30 June, the active region, which we marked in red, was still behind the eastern limb, and by 9 July, it would not have been near the limb yet (Figure~\ref{Fig2}j). Therefore, we conclude that the red sunspot did not become recurrent since its first appearance on 4 June.

On 12 July 1660, Hevelius reported that the spots had clearly disappeared, albeit another smaller spot along with a few tiny ones with umbrae, was observed near the eastern limb. On that day, we calculated that the center of the active region, which we marked in blue, was already beyond the limb. If the spot, which we marked in red (Figure~\ref{Fig2}k), was still alive, it would have been near the edge of the disc, contrary to Hevelius's description. The assumed latitude\,--\,longitude range of a new sunspot group near the eastern limb is shown in green (Figure~\ref{Fig2}k). On 13 July, Hevelius noted that the spots had weakened by the afternoon, and by 19 July, the solar disk was blank.

On 27 July 1660, Hevelius reported the solar disk was absolutely clear. The following day, however, he observed a round spot near the eastern limb, which he claimed had entered the Sun on the day before for the first time. These two notes seem to contradict each other. Figure~\ref{Fig2}l shows the possible location of the sunspot on 27 July (blue); the color fill is removed, as Hevelius noted that no spots were visible on that day. Here and further, we extend the assumed latitude range into the northern hemisphere because Hevelius mentioned that this was the first time he observed the spot, with the hemisphere unspecified. On the next day (Figure~\ref{Fig2}m), only the assumed latitude\,--\,longitude range of the sunspot (blue) matches the observation by Hevelius. If the spot marked in red were still visible, it would have been located at the very edge of the disc on 28 July, which contradicts Hevelius's description.

On 31 July 1660, Hevelius reported that the spot remained near the center of the Sun. On 3 August, the spot had moved into the western quadrant. On 6 August, a much smaller spot appeared near the western limb, and by 7 August, the spot was still visible on the Sun. Figures~\ref{Fig2}n, o, and p show that only the assumed positions of the spot, marked in blue, agree with these observations. Therefore, the sunspot group first observed by Boyle on 4 June 1660 (Figure~\ref{Fig1}c, red arc) was not a long-lived object after all.

In conclusion, we find that from 22 February to 7 August 1660, the object marked in blue (Figure~\ref{Fig2}) was not the same long-lived recurrent sunspot group, but rather an activity nest\footnote{or center of solar activity, activity complex, area of long-term activity, core of activity complex, focus of sunspots, sunspot nestlet, \textit{etc.} depending on the rigour of its definition.} that persisted through several rotations. The first active region, or regions, within this nest were observed over two solar rotations from 22 February to 22 March 1660. A new portion of magnetic flux, emerging as a separate active region, was observed over three rotations from 7 May to 9 July 1660. Another active region, or regions, within this nest lived from 28 July to 7 August 1660. However, its latitudinal position was not specified, so it is possible that this active region belonged to the opposite hemisphere.

The object marked in red (Figure~\ref{Fig2}) was observed for one rotation on 4\,--\,12 June 1660. One more sunspot group, marked in green, was reported on 12 and 13 July 1660. In total, we identified at least five distinct sunspot groups. The sunspot parameters, as derived from our assumption about the drawings, are provided in the Electronic Supplementary Materials.

Modern studies have shown that found that an activity nest can live from 6 up to 15 solar rotations \cite{1986SoPh..105..237C}. \citet{2010SoPh..262..299H} found that only two sunspot nestlets lasted for over nine solar rotations, while five nestlets persisted for eight rotations. The discrepancies in these findings seem to stem from differences in the criteria used to identify a pair of sunspot groups as recurrent. 

\section{Year 1671}
\label{S-1671}

\begin{figure}    
\centerline{\includegraphics[width=1\textwidth,clip=]{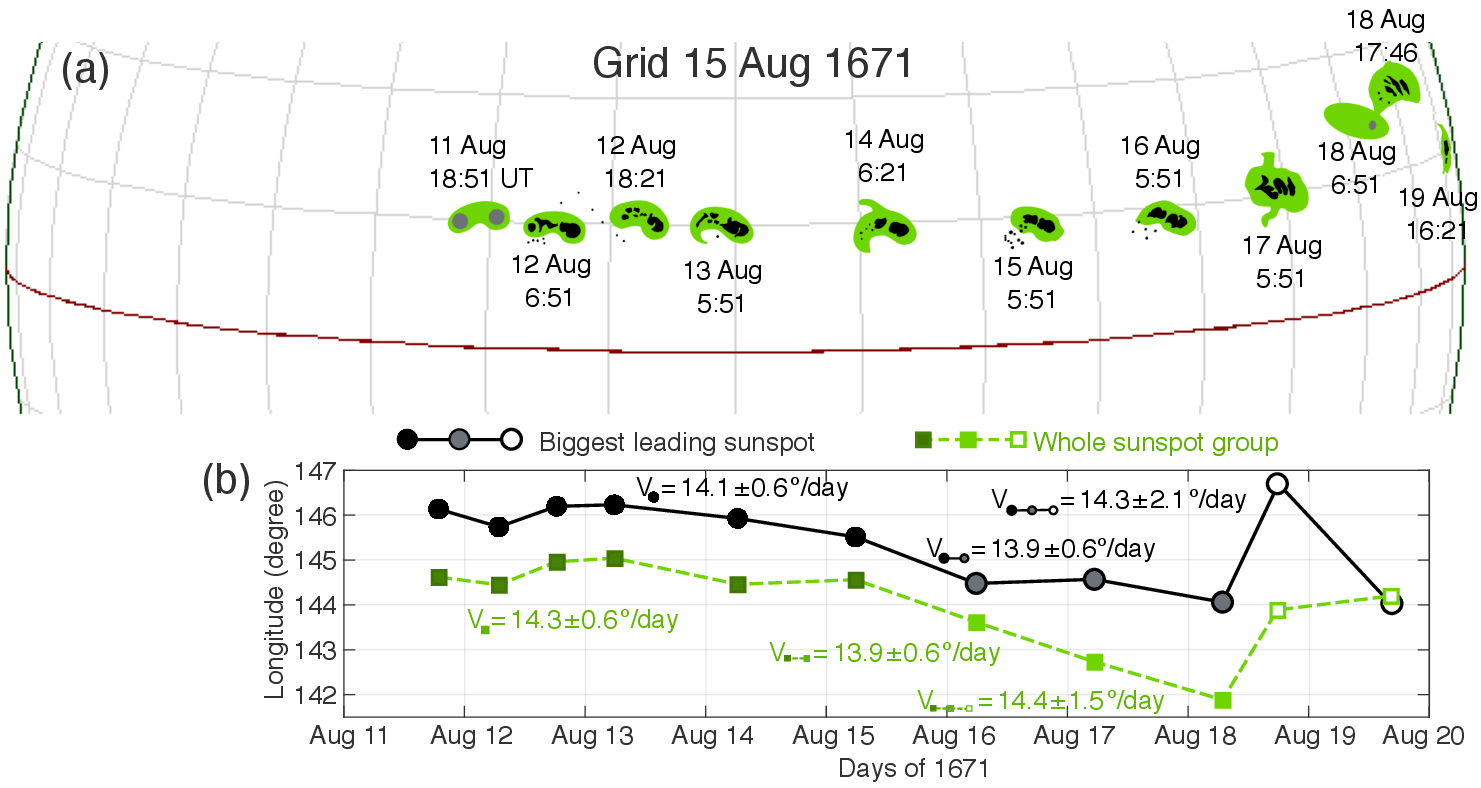}}
\small
        \caption{(\textbf{a}) Sunspot position and size deduced from Cassini's report. The heliographic grid corresponds to the equatorial set-up of a telescope on 15 August 1671. (\textbf{b}) Sunspot group longitudes. \textit{Colored symbols} indicate the data used to calculate the rotation rate.}
\label{Fig3}
\end{figure}

This series of sunspot observations recorded from 11 August to 15 September 1671 was recently analyzed by \citet{2021MNRAS.506..650H}. They got an average latitude of $10\pm 1^{\circ}$ for the observations by Jean-Dominique Cassini on 11\,--\,13 August and $7.5 \pm 2.5^{\circ}$ for Heinrich Siverus on 18 August\,--\,15 September, proposing that both observers documented the same long-lived active region. Below, we expand upon these findings by analyzing sunspot longitudes and rotation rates.

The observations by Jean-Dominique Cassini (also known as Giovanni Domenico Cassini) were published in two French monographs: one covering 11\,--\,13 August \citep{Cassini_1671}, and the other spanning 14\,--\,20 August 1671 \citep{Cassini_1671_2}. On the same year, these monographs were also reprinted in English in the Philosophical Transactions \citep{Cassini_1671_Ph_Tr, Cassini_1671_Ph_Tr_2} allowing a comparison of engraving accuracy between the French and English editions.

Apparently, Cassini used an equatorial-mounted telescope invented by Christoph Grienberger \citep{2003AcHA...18...34D}. For detailed observations of the fine structure of sunspots, Cassini employed a seventeen-foot telescope. A three-foot glass was used for measuring the penumbra's position on the solar disk. His goal was to estimate the rotational velocity of the sunspot, which he stated as the main objective of his research. 

Cassini referred to the penumbra as a ``misty crown" or ``coronet". Only objects sharing this common crown were classified as spots, while tiny objects beyond the crown's boundary were described as ``black points" rather than spots. Cassini noted that some barely visible points were so small that the engraver had to represent them as much larger than they actually were. On the image of the solar disk, the engraving discrepancy in the crown's position is up to $1^{\circ}$, and for the small black points, it is up to $2^{\circ}$. Once, there is a discrepancy in the number of black points recorded in the French and English monographs. Both monographs depict the solar disk as slightly (5\%) vertically elongated, which contrasts with Picard's measurements \citep{Monnier_1741} indicated that the horizontal diameter slightly exceeded the vertical diameter.

Cassini made twelve detailed drawings of the sunspot group, but only the first four were transferred to the solar disk. He recorded the solar semi-diameter ($15'$ $55''$), measured the linear size of the sunspots, and regularly noted the time elapsed between the active region and the solar limb as they crossed the same ``horary circle". This data allows us to map the detailed sunspot drawings onto the solar disk with accuracy up to the orientation of the sunspot group. The uncertainties in Cassini's measurement were approximately one hour for time and a few degrees for sunspot position.

A typical observation was accompanied with the following notation: From six at night to seven, the time between the Sun's center and coronet is one time eight seconds, and another time seven seconds and half. During the initial days, Cassini measured the distance to the center of the spot, then to the inner edge of the penumbra, and finally to the front outer edge of the penumbra.

Figure~\ref{Fig3}a shows our reconstruction of the sunspot group's positions and areas. Sunspot-group areas are, on average, 30\% larger than those estimated from the original solar disk engraving (see Electronic Supplementary Materials). Between 11 and 16 August 1671, the sunspot group follows a latitude of $9.9 \pm 0.5^{\circ}$. However, on 17 and 18 August the latitude increases, reflecting growing uncertainty in the observations.

The observations on 11 August and in the morning of 18 August lack details about the fine structure of the sunspot group; thus, these positions are represented schematically. The final observation on 19 August is a product of suggestion based solely on a brief note indicating that the spot was approximately its own breadth away from the solar limb. Clouds prevented Cassini from taking precise measurements that day.

Figure~\ref{Fig3}b illustrates the derived longitudes. The black points represent the longitude of the largest leading sunspot, while the green points denote longitude of the entire sunspot group. From 15 August 1671 onward, there is a noticeable decline in the longitude values, which likely results from the observational uncertainties. This has a significant impact on the calculated rotation rate.

From 11 to 15 August, the rotation rate is determined to be $14.1 \pm 0.6^{\circ}$ d$^{-1}$ for the largest leading sunspot and $14.3 \pm 0.6^{\circ}$ d$^{-1}$ for the entire group. When including data up to 18 August, the rate for the largest spot decreases to $13.9 \pm 0.6^{\circ}$ d$^{-1}$. For the complete set of data, the rotation rates are $14.3 \pm 2.1^{\circ}$ d$^{-1}$ for the largest spot and $14.4 \pm 1.5^{\circ}$ d$^{-1}$ for the entire group.

\begin{figure}    
\centerline{\includegraphics[width=1\textwidth,clip=]{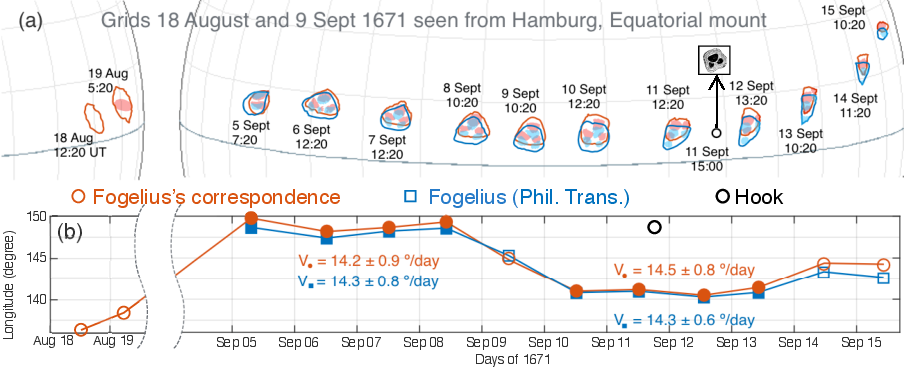}}
\small
        \caption{(\textbf{a}) Superposition of the colored modifications of the original drawings by Siverus from \citet{Fogelius_1671} is shown in \textit{red} and from \citet{Fogelius_1671_Ph_Tr} in \textit{blue} with imposed heliographic grids. Drawing by \citet{Hook_1671} is shown in \textit{black}. (\textbf{b}) Sunspot-group longitudes. \textit{Filled symbols} indicate the data used to calculate the rotation rate.}
\label{Fig4}
\end{figure}

The series of observations by Heinrich Siverus accompanies a letter by Martin \citet{Fogelius_1671} to Henry Oldenburg, which is preserved in the Royal Society Library. Portion of these observations was also reprinted in \citet{Fogelius_1671_Ph_Tr}. Figure~\ref{Fig4}(a) illustrates the superposition of the Siverus's drawings, represented in red and blue. The heliographic grid overlaid on the image corresponds to the equatorial mounting of the telescope used. Compared to Cassini's engravings, Siverus's depictions are more schematic, with sunspots appearing enlarged.

The engraving discrepancies are approximately $2^{\circ}$ in both latitude and longitude. Similar to the observations by \citet{Cassini_1671_2}, the latitude of the sunspot group increases as the group approaches the western limb, likely resembling increasing observational uncertainty. The latitude range of the sunspot group is estimated to be $2-10^{\circ}$, with an average latitude of $7.3 \pm 3.4^{\circ}$.

Figure~\ref{Fig4}(b) displays the longitude variations of the sunspot group, which exhibit considerable fluctuation. The rotation rate is estimated to be $14.2 \pm 0.8^{\circ}$ d$^{-1}$ and $14.5 \pm 0.8^{\circ}$ d$^{-1}$ for two distinct intervals, as indicated by the filled symbols. The actual value is likely in the range of 14.2\,--\,$14.3^{\circ}$ d$^{-1}$, aligning well with Cassini's measurements.

Another sunspot observation by Robert Hook (commonly spelled Hooke) is represented as a black circle in Figure~\ref{Fig4}. Similar to Cassini, \citet{Hook_1671} referred to the sunspot as comprising the umbra alone, describing the penumbra as a ``dusky cloud". He also provided a sketch of the sunspot group (enclosed in a black frame in Figure~\ref{Fig4}a) and noted that it was ``of this form exactly". In addition to the sketch, Hook recorded the time, size, and position of this group. The black circle in Figure~\ref{Fig4} depicts the position and size of the whole phenomenon based on Hook's measurements, as it would appear from Hamburg (where Siverus observed; Hook himself was in London). The size of the whole phenomenon (1/72 of the solar diameter) is evidently too small when compared to Siverus's drawings. However, the longitude of the sunspot group reported by Hook aligns closely with the first part of Siverus's observations, as illustrated in Figure~\ref{Fig4}b.

By examining the first longitudes reported by Cassini ($\approx 145^{\circ}$ on 11\,--\,14 August), those by Siverus ($\approx 148^{\circ}$ on 5\,--\,8 September), and the one by Hook ($148^{\circ}$ on 11 September 1671), we infer that the group was likely recurrent, with a rotational rate of approximately $14.3^{\circ}$ d$^{-1}$. However, it is noteworthy that Siverus generally mapped the sunspot a few degrees closer to the Equator compared to Cassini. This positinal difference raises the possibility that the spots observed in August and September may have belonged to two distinct groups within the same activity nest.

\section{Year 1672}
\label{S-1672}

\begin{figure}    
\centerline{\includegraphics[width=1\textwidth,clip=]{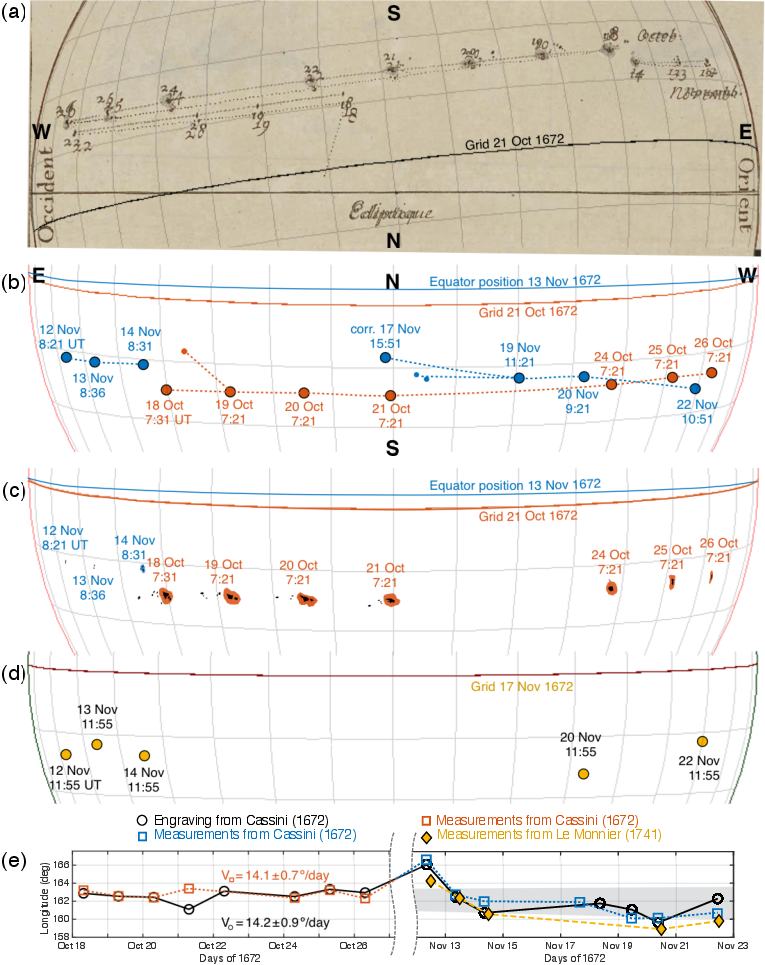}}
\small
        \caption{(\textbf{a}) Superposition of two original engravings by \citet{Cassini_1672} and \citet{Bion_1751}. The heliographic grid corresponds to 21 October 1672 with \textit{bold letters} marking the cardinal points. (\textbf{b}) Largest sunspot positions based on tabular measurements: observations from 18\,--\,26 October 1672 are shown in \textit{red}, and those from 12\,--\,22 November 1672 in \textit{blue}. The heliographic grid corresponds to 21 October, and the Equator position on 13 November marked in \textit{blue}. \textit{Small circles} highlight discrepancies in sunspot mapping. (\textbf{c}) Sunspot positions and sizes as deduced by us: observations from 18\,--\,26 October are shown in \textit{red} and those from 12\,--\,14 November 1672 in \textit{blue}. (\textbf{d}) Sunspot position derived from tabular measurements in \citet{Monnier_1741}. (\textbf{e}) Sunspot longitudes derived from the engraving are represented in \textit{black} and those from the tabular measurements are in \textit{color} with the corresponding rotation rates. The \textit{gray shaded area} illustrates the expected range of the longitudes in November.}
\label{Fig5}
\end{figure}

The series of observations initiated by Ole Christensen Roemer (Rømer) and continued in collaboration with Jean Picard at the Paris observatory from 18 October to 22 November 1672 was published by \citet{Cassini_1672} and \citet[reprinted from 1699]{Bion_1751}. These publications include engravings, text, and conclusions, though they differ in some aspects.

Tabular measurements of the largest sunspot's position are available only in \citet{Cassini_1672}. Figure~\ref{Fig5}(a) presents a superposition of two original engravings from \citet{Cassini_1672} and \citet{Bion_1751}. The engraving discrepancy in latitude and longitude ranges from $1-2^{\circ}$. Detailed sunspot parameters reconstructed from both engravings are provided in The Electronic Supplementary Materials.

Further, the text description is derived from \citet{Cassini_1672}. The author referred to the umbra as the sunspot and described the penumbra as a ``cloud". Observations were interrupted by bad weather on 26 October; however, Cassini, observing from Provence, 
recorded seeing the sunspot the following day at Noon as it touched the western limb. He noted that the sunspot appeared narrower toward the limb due to the projection effect. This apparent reduction led to the conclusion that the spot might reappear on the eastern limb.

By 9 November, the spot had not yet reappeared, and cloudy conditions prevailed the following day. Observations resumed on 12 November. Poor weather hindered observations until 18 November, when the sunspot was visible again as a small black dot. At least six-foot telescope was required to observe it. These measurements were employed to determine the rotation rate of the Sun, the tilt of the solar rotation axis, and the regularity of sunspot motion (see Section~\ref{S-Oct_1676}). \citeauthor{Cassini_1672} concluded that the object observed over two months was the same recurrent sunspot, continuously located $15^{\circ}$ south of the Equator. He attributed the divergence in sunspot tracks in the October and November engravings to the annual variation in the solar axis tilt relative to the observer. Based on Cassini's assumption of a long-lived active region, we evaluate its average latitude to be $-13.4 \pm 0.2^{\circ}$ in \citet{Bion_1751} and $-14.1 \pm 0.1^{\circ}$ in \citet{Cassini_1672}.

Figure~\ref{Fig5}(b) illustrates the positions of the largest sunspot based on tabular measurements: October observations are marked in red, and November observations in blue. Measurement on 22 October is absent. The heliographic grid corresponds to 21 October (in red), and the Equator position on 13 November 1672 is marked in blue. It is suggested that Roemer and Picard used the equatorial setup of a telescope for these observations. The distance from the western limb was given in time units, while that from the South was measured in angular units. The ratio of the solar diameter measured in these units decreases by 5\% over the observation period, which was accounted for in transferring the tabular measurements onto the solar disk.

On 18 October 1672, the recorded distance of the sunspot from the South is given as $9'$ $5''$ (small red circle in Figure~\ref{Fig5}b). We suspect this to be a misprint and have assigned a corrected value of $7'$ $5''$ (regular red circle). The corrected data for this day is provided in the Electronic Supplementary Materials.

On 18 November at 7 a.m. local time, the measured distance from the western limb was recorded as $1'$ $5''$ (regular blue circle). This measurement significantly differs from the sunspot position shown in the engravings indicated by two small blue circles). It seems that either author or publisher of the original observations proposed that the measured distance was incorrect, leading to a corrected sunspot mapping in the engraving. In contrast, we intentionally hypothesize that the measured distance was accurate, but a time error occurred. We suggest the correct time might have been 17 November 15:51~UT. The corrected data based on this assumption is included in the Electronic Supplementary Materials. Overall, the measurements yield an average sunspot latitude of $-14.6 \pm 1.5^{\circ}$.

Roemer and Picard also produced ten detailed drawings of the observed sunspots and provided measurements of the penumbra's linear size. These detailed drawings allow us to transfer the sunspot positions onto the solar disk with accuracy limited by the orientation of the active region (Figure~\ref{Fig5}c): October observations are in red and November observations in blue. 

Figure~\ref{Fig5}(d) presents the sunspot position measured by Picard, as published by \citet{Monnier_1741}, later reprinted by \citet{La_Lande_1778}, and incorporated by \citet{Spoerer1889}. Spoerer evaluated the sunspot latitude as $-13^{\circ}$, while our analysis, based on the solar altitude measurements near Noon, determined an average latitude of $-13.7 \pm 1.8^{\circ}$. However, two November observations present in \citet{Cassini_1672} are missing in \citet{Monnier_1741}.

Figure~\ref{Fig5}(e) shows the sunspot longitudes of the largest sunspot derived from engraving (black) and measurements (red and blue) as reported by \citet{Cassini_1672}. Based on the most reliable October observations, the engraving indicates a sidereal rotation rate of $14.2 \pm 0.9^{\circ}$ d$^{-1}$ ($13.3 \pm 0.9^{\circ}$ d$^{-1}$ in synodic units), while the measurements suggest $14.1 \pm 0.7^{\circ}$ d$^{-1}$ ($13.1 \pm 0.7^{\circ}$ d$^{-1}$ in synodic units). The gray shaded area indicates the potential range of longitudes for the active region in November, assuming a motion velocity of $14.1-14.2^{\circ}$ d$^{-1}$. The November data (black, blue, and yellow) shows greater scatter, but are shifted toward $14.1^{\circ}$ d$^{-1}$. Thus, matching longitudes confirms that the observers likely saw an active region spanning two solar rotations.

\citet{2006SoPh..234..379C} analyzed the engraving from the second edition of Bion's book and obtained significantly different results for observation dates, sunspot latitudes, and synodic rotation rate. Their findings highlight that relying solely on the engraving, which contains incomplete information, may lead to the conclusion that the solar equatorial rotation during the Maunder minimum was slower.

A rough translation of the report accompanying the measurements reads as follows: on 12 November 1672, Monsieur Picard and Monsieur Romer, while at the Royal Observatory, discovered a sunspot resembling an ant. On 13 November, the spot split into two parts, and on the 14th, a cloud (penumbra) appeared around it, on the edge of which appeared a third spot. On 18 November, they saw the spot as a small black dot. On 22 November, observing the spot required at least a six-foot telescope. It was surrounded by small intermittent clouds and faculae.

It seems that on 26 October, the sunspot retained its penumbra, whereas on 12 and 13 November, it appeared only as a pore (without penumbra) measuring 8~msh, as shown in both engravings and our reconstruction (Figure~\ref{Fig5}d). The penumbra reported on 14 November implies the emergence of a new sunspot, reinforcing the hypothesis that the observers were witnessing an activity nest, rather than a single, long-lived sunspot group.

\section{June 1676}
\label{S-1676}

\begin{figure}    
\centerline{\includegraphics[width=1\textwidth,clip=]{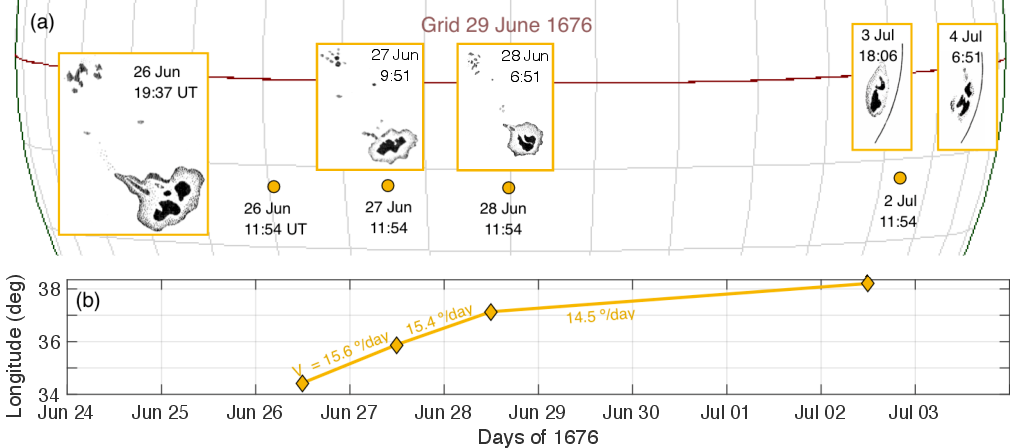}}
\small
        \caption{(\textbf{a}) Sunspot position derived from the tabular measurements in \citet{Monnier_1741}. A few horizontally and vertically mirrored sunspot reproduced from the original engravings are highlighted in \textit{yellow frames}. (\textbf{b}) Longitudes and rotation rate determined in this study.}
\label{Fig6}
\end{figure}

An active region observed in June\,--\,July 1676 was described by \citet{Hook_1677} as ``a very conspicuous Macula with its immediatly incompassing Nubecula [interpreted as a cloud, often used to describe a penumbra] and some other less conspicuous Spots at a further distance pass over the Disk of the Sun". Hook hypothesized that these sunspots were the cause of the extraordinary heat experienced in England and Europe during that period.

A portion of the sunspot position measurements, presumably made by Picard near Noon using a quadrant with a radius of 32 inches from 26 June to 2 July 1676, was published by \citet{Monnier_1741}. Originally the last measurement was listed as 1 July, but upon matching the sunspot longitudes, we suspect this is a typographical error. A detailed engraving of this sunspot group spans from 26 June to 4 July. Most of the drawings were made in the morning or evening, when the Sun's altitude is low. A few reproductions of these drawings are shown in yellow frames in Figure~\ref{Fig6}(a).

Hypothesizing that Picard exploited the same equipment as \citet{Cassini_1671, Cassini_1671_2}, we flipped the engraving both top to bottom and right to left, resulting in an anti-Joy orientation of the sunspot group. Additionally, the trailing spots may represent a different group, located approximately 5\,--\,$8^{\circ}$ closer to the Equator than the large leading spot. Since Picard did not measure the size of these sunspots, we cannot transfer the engraving directly to the solar disk. In the Electronic Supplementary Materials, we provide the count of umbras and pores based on the engraving.

The solar diameter must be known to determine sunspot coordinates from angular distance measurements. We interpolated the routine measurements by Picard from the period 1666\,--\,1670 and determined the angular size of the Sun's disc to be $31'$ $37.5''$. From this, we obtain a sunspot latitude of $-12.4 \pm 0.2^{\circ}$. Measurements published in \citet{Monnier_1741} later reprinted by \citet{La_Lande_1778}, and \citet{Spoerer1889}, reported a latitude of $-13^{\circ}$.

Figure~\ref{Fig6}(b) shows the derived longitudes and rotation rate calculated between consecutive observations. Due to the limited number of observations and the scatter of longitudes, the rotation rate is difficult to judge. The longitude increased by $3.8^{\circ}$ over six days, which we suspect is an artifact. This apparent growth in longitude could be due to several factors: (i) variations in sunspot position caused by changes in sunspot shape (referred to as \textit{La tache} in French, likely the largest umbra), and (ii) imperfections in the pendulum time measurements. An error of one second in time measurements results in an inaccuracy of 1\,--\,$1.5^{\circ}$ in the sunspot's position at the center of the solar disk, and about $3^{\circ}$ for a spot located $60^{\circ}$ from the center. \citet{La_Lande_1778} also questioned the accuracy of the sunspot position and solar altitude measurements at the Paris observatory, citing the lack of a micrometer and vernier and the irregularity of the sunspot shape. In contrast, \citet{1993A&A...276..549R} argued that the Paris observatory was equipped with a micrometer.

\section{August 1676}
\label{S-Aug_1676}

\begin{figure}    
\centerline{\includegraphics[width=0.5\textwidth,clip=]{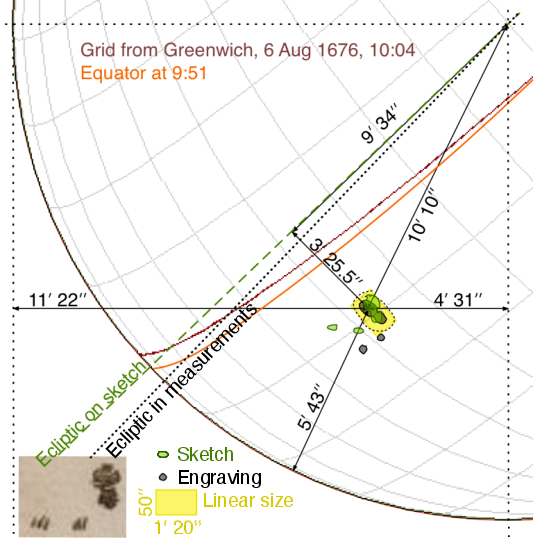}}
\small
        \caption{Sunspot group position as derived from the original sketch by \citep{Flamsteed_1676} in \textit{green} and that from the combined results from measurements \citep{Flamsteed_1676_Hist} and engraving \citep{Flamsteed_1676_Ph_Tr} in \textit{black}. The heliographic grid corresponds to 6 August 1676 at 10:04 in Greenwich, with the Equator position as observed at 9:51 highlighted in \textit{orange}. The Ecliptic from the original sketch represented by the \textit{dashed green line}, while the \textit{dotted black line} shows the Ecliptic based on the measurements. The \textit{yellow rectangle} marks the linear size of the sunspot group. The original image of the sunspot is displayed in the bottom-left corner.}
\label{Fig7}
\end{figure}

Analyzing the sunspot's rotation, \citet{Cassini_1676_Ph_Tr} concluded that the sunspot (\textit{Macula}) observed in August and the one from late June were distinct objects. He noted that the late-June sunspot was located farther from the Equator than the August sunspot. \citet{Hook_1677} also observed this active region and wrote that, on 8 August 1676, he counted ``about six greater and smaller [spots] in one knot with their proper Nubecules".

Sunspot observations made by John Flamsteed in August 1676 (Gregorian calendar; the original dates correspond to the Julian calendar) were published in \citet{Flamsteed_1676_Hist} as a table of detailed measurements obtained using a micrometer. However, this table contains several misprints. A shortened version of Flamsteed's measurements, along with those by Edmond Halley in Oxford, was included in \citet{Flamsteed_1676_Ph_Tr} and accompanied by the engraving. Additionally, Flamsteed's measurements on 6 August 1676 with a sunspot sketch are included in a letter to Jonas Moore \citep{Flamsteed_1676}. The transcription of this text can be found in \citet{2016SoPh..291.2493C}. \citet{Flamsteed_1676} described the spot as significantly large, appearing slightly divided in the middle, with two thin, cloudy spots following it. Although, the sunspot was understood to refer to its umbra, the linear dimensions provided for the large sunspot correspond its penumbra (highlighted by a yellow rectangle; the original image is located at the bottom left of Figure~\ref{Fig7})).

Flamsteed used an alt-azimuth mount for his eight-foot telescope, which was occasionally shaken by the wind. He took the measurements over periods ranging from a few minutes to an hour, introducing uncertainty into the mapping of sunspot. An example of this is shown in Figure~\ref{Fig7}. During the observation, which lasted from 9:51 to 10:04, the change in the Equator's position is represented in different colors. The reversed and flipped reproduction of the sunspot group and the Ecliptic from Flamsteed's original sketch \citep{Flamsteed_1676} is shown in green, while the measurements are depicted in black. Additionally, we have overlaid the sunspot group from the engraving \citep{Flamsteed_1676_Ph_Tr} in black. To reconcile all the measurements, the Ecliptic from the sketch was used. Relying on the measured Ecliptic, the sunspot group is placed $1.5^{\circ}$ farther from the Equator, which is inconsistent with other measurements. These cumulative uncertainties affect the final determination of the sunspot's position.

\begin{figure}    
\centerline{\includegraphics[width=1\textwidth,clip=]{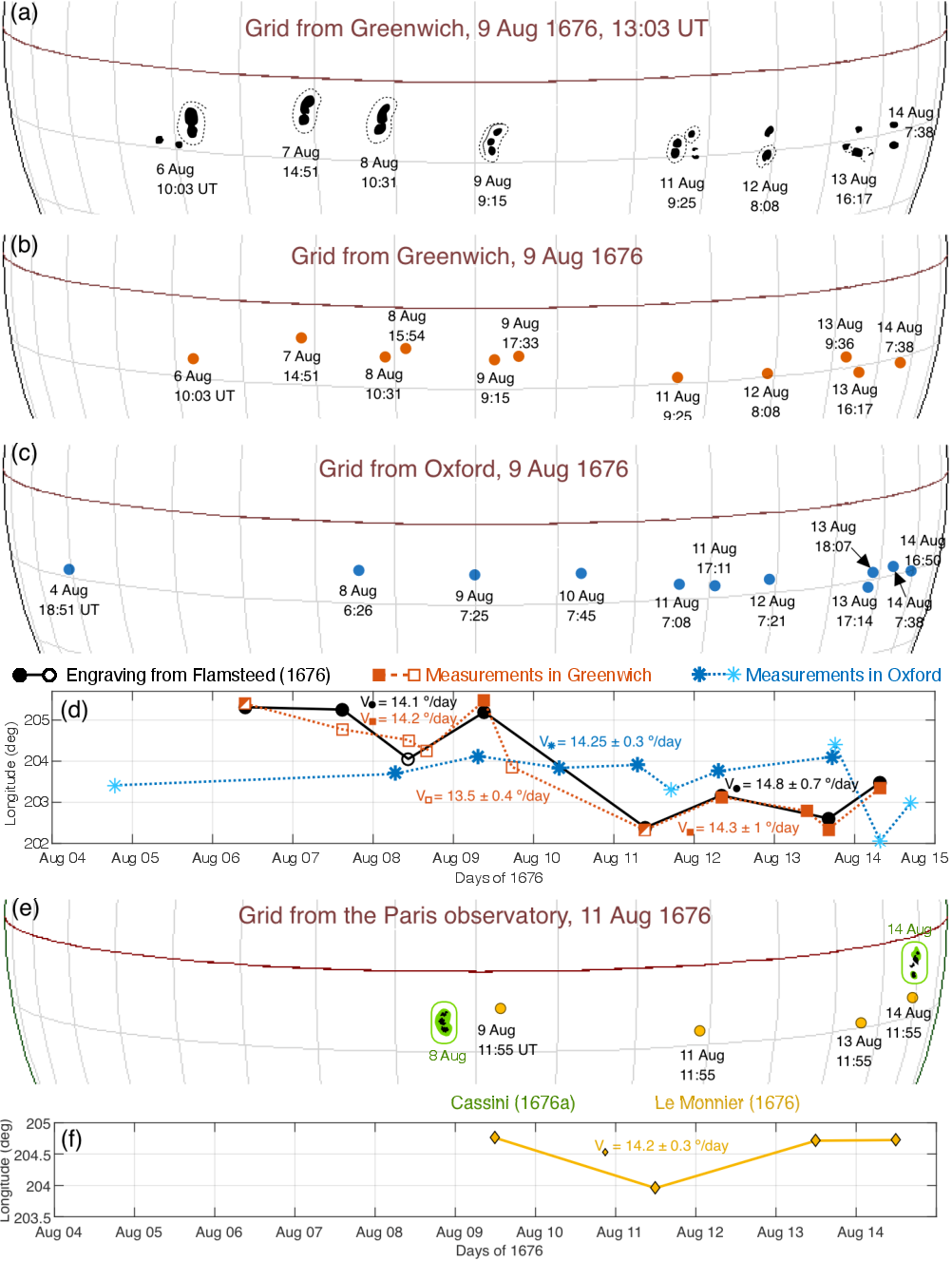}}
\small
        \caption{(\textbf{a}) Reproduction of the original engraving from \citet{Flamsteed_1676_Ph_Tr}. (\textbf{b}) Sunspot position derived from tabular measurements in Greenwich, with \textit{filled circles} representing reliable data, and \textit{unfilled circles} indicating unreliable. (\textbf{c}) Sunspot positions from measurements in Oxford. (\textbf{d}) Sunspot longitudes from the engraving (\textit{black}), tabular measurements in Greenwich (\textit{red}), and Oxford (\textit{blue}). Rotation rates are derived based on reliable data indicated by \textit{filled symbols}. (\textbf{e}) Sunspot position from measurements at the Paris Observatory \citep{Monnier_1741} shown in \textit{yellow}. Drawings from \citet{Cassini_1676_Ph_Tr, Cassini_1676} are embedded in \textit{green}. (\textbf{f}) Sunspot longitudes and the corresponding rotation rate.}
\label{Fig8}
\end{figure}

Figure~\ref{Fig8}(a) reproduces the original engraving \citep{Flamsteed_1676_Ph_Tr} with an imposed heliographic grid. Corrections were made for the  asymmetry of the solar disk caused by uneven sizes of vertical and horizontal diameters, which differed by up to 6\%. On 6 August 1676, the penumbra line appears broken and does not fully close. Flamsteed mentioned observing a crack, confirming that this is not merely an engraving inaccuracy. On 13 August, three sunspots (\textit{Maculae tres}) were observed, with the southernmost one having a thin penumbra (\textit{tenuis nubeculisa}). The latitude of the sunspot group ranged from $-4^{\circ}$ to $-10^{\circ}$, with an average of $-7.9 \pm 1.1^{\circ}$. \citet{Spoerer1889} later estimated the latitude to be approximately $-6^{\circ}$.

In measurements by \citet{Flamsteed_1676_Hist}, the solar diameter varied from $31'$ $46''$ to $31'$ $55''$, but it was not provided for every observation, introducing some uncertainty. Note that routine measurements at the Paris observatory from 6 to 14 August 1666\,--\,1670 showed a gradual increase in the solar diameter from $31'$ $45''$ to $31'$ $48.5''$. For the observation on 12 August, we corrected a typo in the recorded distance to the center of the solar disc ($9'$ $49''$ instead of $9'$ $19''$). Sunspot positions were reconstructed for each day by adjusting the image sizes and aligning them with the solar rotation axis, as shown in Figure~\ref{Fig8}(b). The results reveal uncertainty in sunspot position indicated by the red-filled and gray unfilled circles. These uncertainties arose from measurements that sometimes spanned up to an hour. Heliographic coordinates were determined based on the more reliable positions represented by the filled symbols, yielding an average latitude of $-7.1 \pm 1.4^{\circ}$.

Figure~\ref{Fig8}(c) illustrates the sunspot position derived from Edmond Halley's measurements in Oxford. \citet{Flamsteed_1676_Ph_Tr} provided a brief table of observations and, therefore, we aligned the sunspots with the Ecliptic. The average latitude obtained from these measurements is $-8.3 \pm 1.5^{\circ}$. On 15 August, Halley observed the sunspot at 8:35~UT when it was near the limb. However, due to projection shortening and the high altitude of the Sun, he was unable to take accurate measurement. Both Flamsteed and Halley measured the distance to the center of the largest sunspot (\textit{media Macula}) or the midpoint of sunspots (\textit{Maculae medium}), which also introduced some degree of uncertainty in sunspot mapping.

Figure~\ref{Fig8}(d) shows the longitude derived from the engraving (sunspot number 2 in the Electronic Supplementary Materials) and the measurements. Observations by Flamsteed, represented by black and red symbols, exhibit a consistent pattern: larger longitudes before 10 August 1676 and smaller longitudes afterward. We attribute this pattern to inherent observational uncertainties, as similarly noted in Figures~\ref{Fig3} and \ref{Fig4}. Longitude variations based on the engraving (black points) indicate rotation rates of $14.2 \pm 0.7^{\circ}$ d$^{-1}$ and $14.8 \pm 0.7^{\circ}$ d$^{-1}$, before and after 10 August 1676, respectively, marking these results as unreliable. The measurements (red filled symbols) suggest rotation rates of $14 \pm 0.6^{\circ}$ d$^{-1}$ and $14.2 \pm 1^{\circ}$ d$^{-1}$ for the same periods. Additionally, Figure~\ref{Fig8}(d) includes measurements from Oxford, indicated by blue stars. To determine the rotation rate, we utilized only reliable sunspot positions (dark blue symbols) and calculated a rate of $14.25 \pm 0.3^{\circ}$ d$^{-1}$. 

Figure~\ref{Fig8}(e) illustrates the position of the sunspot (\textit{la tache}) as retrieved from the measurements by Picard and/or Cassini at the Paris observatory using quadrants with radii of 3 foot and 32 inches, recorded from 9 to 14 August 1676 and later published by \citet{Monnier_1741}. Additionally, we embed sunspot drawings from \citet{Cassini_1676_Ph_Tr, Cassini_1676} in green. The sizes of these spots are chosen arbitrarily, and the observation times were not noted. The latitudinal position of the sunspots ($3'$ from the disk center, as stated by Cassini) align well with the measurements of Flamsteed and Halley.

To derive the heliographic coordinates of the sunspot, we assume the angular size of the solar disc to be $31'$ $46''$ based on Picard's measurements. The resulting latitudes range from $-4.6$ to $-9.6^{\circ}$, with an average of $-7.6 \pm 1.8^{\circ}$. Figure~\ref{Fig8}(f) shows the corresponding longitudes with a derived rotation rate of $14.2 \pm 0.3^{\circ}$ d$^{-1}$, which aligns with Flamsteed's and Halley's measurements. This result is considered more reliable due to small scatter in longitudes (less than $1^{\circ}$).

\section{October\,--\,December 1676}
\label{S-Oct_1676}

\begin{figure}    
\centerline{\includegraphics[width=1\textwidth,clip=]{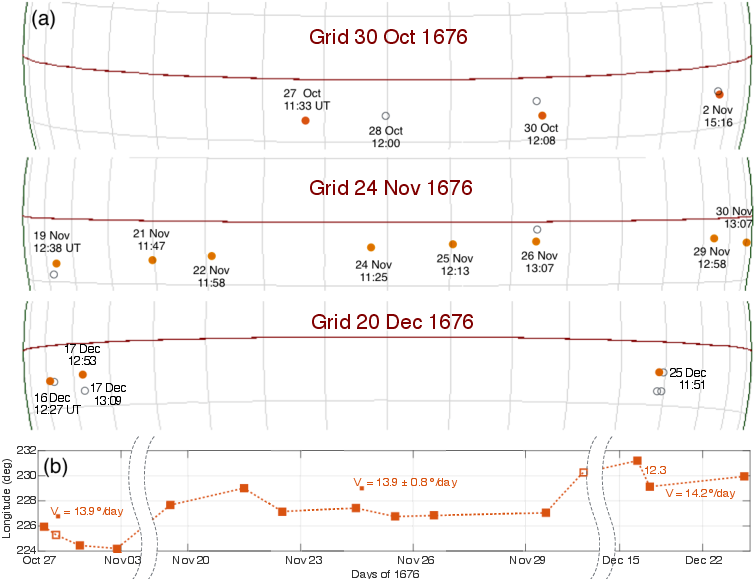}}
\small
        \caption{(\textbf{a}) Sunspot position derived from tabular measurements in \citet{Flamsteed_1676_Hist}. Reliable data are represented by \textit{filled circles}, while unreliable data are shown as \textit{unfilled circles}. (\textbf{b}) Sunspot longitudes and rotation rate deduced based on reliable data.}
\label{Fig9}
\end{figure}

In October\,--\,December 1676, there were several series of sunspot observations. On 27 October 1676 (17 October in the Julian calendar), \citet{Flamsteed_1676_Hist} noted that D. Haynesius (presumably Domino Edvardo Haynesio, or Sir Edward Haynes) visited him and reported observing a sunspot. Together, they used a 16-foot tube to study its shape, as illustrated in the referenced figure. Flamsteed measured its position using a shorter 8-foot tube. Despite locating several copies of \textit{Historia Coelestis Britannica, volumen primum}, page 367 --- which contains the referenced figure --- was missing. In a letter to Towneley, Flamsteed mentioned that the sunspot appeared on 24 or 25 October \citep{2016SoPh..291.2493C}.

Adverse weather conditions on 28 and 29 October interrupted the observations. On 2 November, Flamsteed reported that careful observations made on 27 and 30 October indicated that the \textit{Macula} would have been near the central meridian at Noon on 28 October. He further predicted that the sunspot might reappear again on the limb on 18 or 19 November and return to the central meridian by 24 November.

On 19 November, thick ``vapours" allowed Flamsteed to observe the Sun with the naked eye. Using a longer tube without the red glass typically employed to protect the eyes, he clearly observed the sunspot, as depicted in figure (apparently on page 367). Flamsteed noted that the penumbra (\textit{nubecula}) appeared completely elliptical and expressed surprise that it was significantly wider on the limb side, while the sunspots (\textit{Maculae}) near the disc center seemed almost adjacent (\textit{cohaerere}). Remarkably, Flamsteed described a projection effect nowadays known as the Wilson effect, nearly a century before it was formally identified.

By 30 November, the \textit{Macula} appeared thin due to its proximity to the limb, making it challenging for Flamsteed to accurately measure its position. On 16 December, before the sunspot dissappeared from the visible disk, Flamsteed observed that it had a consistency (\textit{consistentiam}) suggesting that it might persist into the next rotation. On this 16th day, using a shorter tube, he recorded the sunspot's return.

From 17 to 25 December, the weather was extremely cold, with heavy snow and sky continuously obscured by clouds or a dense fog. On 25 December, around Noon, the clouds began to clear, revealing that the \textit{Macula} was still visible, though it had diminished in size and was located near the limb. Its diameter was approximately $15''$, and certainly no larger than $22''$, likely representing the size of the umbra.

Figure~\ref{Fig9}(a) shows the sunspot position from 27 October to 25 December 1676, reconstructed from the tabular measurements in \citet{Flamsteed_1676_Hist}. The sunspot position on 28 October (represented by an unfilled circle) is our interpretation informed by the textual description. During these months, Flamsteed typically made measurements over 15\,--\,30 minutes around Noon. Since the parallactic angle changes by $5^{\circ}$ around Noon, the position of the sunspot is uncertain, as indicated by the red-filled and gray unfilled circles. Heliographic coordinates are determined using more reliable positions marked by filled symbols. The size of the circles is $22''$, as Flamsteed noted on 25 December. The solar diameter varied from $32'$ $28''$ to $32'$ $46''$ and was not given for every measurement, introducing additional uncertainty. We estimate the accuracy of the solar diameter measurement to be about $10''$, which impacts the accuracy of sunspot mapping near the limb. The average latitude is $-5.3 \pm 1.4^{\circ}$. 

Figure~\ref{Fig9}(b) illustrates the derived longitudes. We divided the data into three time intervals to estimate the sunspot rotation rate. In October, we exclude the unfilled symbol on 28 October, yielding a rotation rate of $13.9^{\circ}$ d$^{-1}$. However, this result, along with the sunspot from December, seems unreliable due to the limited number of measurements. In November, excluding the questionable sunspot position (unfilled symbol) on 30 November, we obtain a rotation rate of $13.9 \pm 0.8^{\circ}$ d$^{-1}$.

\begin{figure}    
\centerline{\includegraphics[width=1\textwidth,clip=]{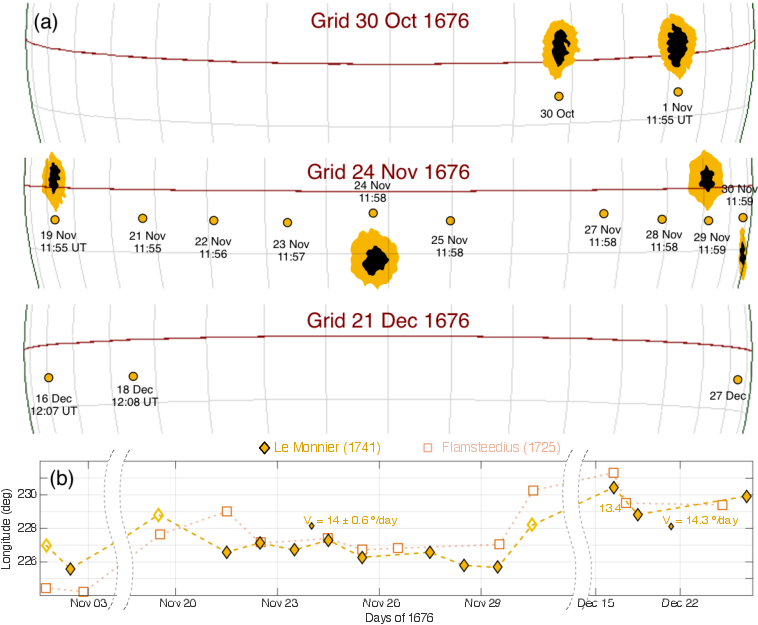}}
\small
        \caption{(\textbf{a}) Sunspot position derived from the tabular measurements in \citet{Monnier_1741} with the horizontally and vertically mirrored sunspot reproductions. (\textbf{b}) Sunspot longitudes and rotation rate. Longitudes from \citet{Flamsteed_1676_Hist} included for comparison.}
\label{Fig10}
\end{figure}

Astronomers at the Paris Observatory, presumably Picard, believed they observed the same long-lived recurrent sunspot. On 19 November 1676, Picard wrote about the return of the sunspot, which had appeared on 30 October and 1 November. On 15 December, he noted that at Noon the sunspot returned for the second time; it almost reached the eastern edge of the Sun and could only be observed with a 20-foot telescope.

Figure~\ref{Fig10}(a) depicts the sunspot's position from 30 October to 27 December 1676, reconstructed from the tabular measurements in \citet{Monnier_1741}. These measurements were typically taken around Noon and are accompanied by sunspot engravings. We include horizontally and vertically mirrored reproductions of these engravings. We suggest that the detailed drawings were made during the morning or evening hours (similar to Figure~\ref{Fig6}) and that the sunspot may have had small trailing spots that were not considered as sunspots \citep{Cassini_1671_Ph_Tr, Cassini_1671_Ph_Tr_2}.

To restore sunspot coordinates, we interpolated the solar diameter based on Picard's measurements: $32'$ $25''$ in October, $32'$ $37''$ in November, and $32'$ $42''$ in December 1676. The time of the solar disc's passage through the Meridian also varied and was taken into account. Our estimate for the average latitude, $-5 \pm 0.7^{\circ}$, aligns with \citet{Spoerer1889} table, who reported $-5.2^{\circ}$ for the latitude in October, $-4.6^{\circ}$ in November, and $-4.9^{\circ}$ in December 1676, based on the Parisian measurements reprinted by \citet{La_Lande_1778}. The longitudes of the sunspot are shown in Figure~\ref{Fig10}(b) as yellow diamonds.

The exact observation time on the first day is unknown; we assumed it was Noon to derive heliographic coordinates. Picard wrote: ``A spot appeared on the Sun, which due to bad weather could not be observed earlier". This implies the observation may have occurred a few hours later, which would reduce the calculated longitude of the sunspot. 

In November, excluding the near-limb observations, we derive a rotation rate of $14 \pm 0.6^{\circ}$ d$^{-1}$. December observations are too sparse to determine the rotation rate. For comparison, in Figure~\ref{Fig10}(b) also includes the longitudes derived from \citet{Flamsteed_1676_Hist}, indicated by pink squares. Both the French and English reports yield that in November, the rotation rate was slower than the sidereal Carrington rotation of $14.18^{\circ}$ d$^{-1}$. Consequently, after a half-turn on the Sun's far side, the longitude of the sunspot would have decreased by $2.3^{\circ}$ at $14^{\circ}$ d$^{-1}$ or by $3.6^{\circ}$ at $13.9^{\circ}$ d$^{-1}$.

However, at each reappearance of the sunspot at the eastern limb, its longitude slightly increases (Figure~\ref{Fig10}). In conclusion, we we found no evidence to support the idea that the active regions observed from October to December 1676 were the same long-lived recurrent sunspot group. It is important to note that the derived longitudes and rotation rates are compromised by numerous uncertainties.

\begin{figure}    
\centerline{\includegraphics[width=0.7\textwidth,clip=]{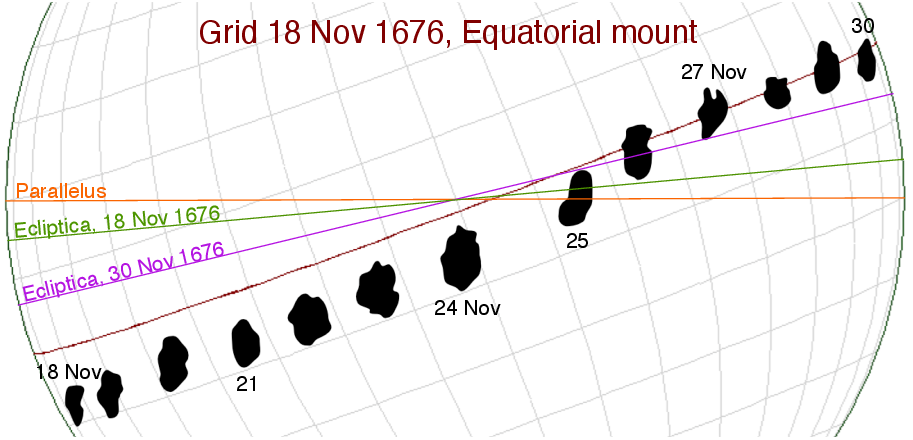}}
\small
        \caption{Superposition of the color-enhanced modification of the original engraving from \citet{Cassini_1730} with the heliographic grid corresponding to the equatorial mount of a telescope.}
\label{Fig11}
\end{figure}

Another astronomer who documented the October\,--\,December sunspot was Cassini. Figure~\ref{Fig11} reprints the colored modification of the original engraving from \citet{Cassini_1730} with an overlaid heliographical grid. The image size is small, while the sunspots themselves are enlarged. Consequently, the engraving is coarse and not suitable for calculations. However, it serves as a useful illustration of the uncertainties inherent in such engravings. Further, we process more accurate engravings.

\begin{figure}    
\centerline{\includegraphics[width=1\textwidth,clip=]{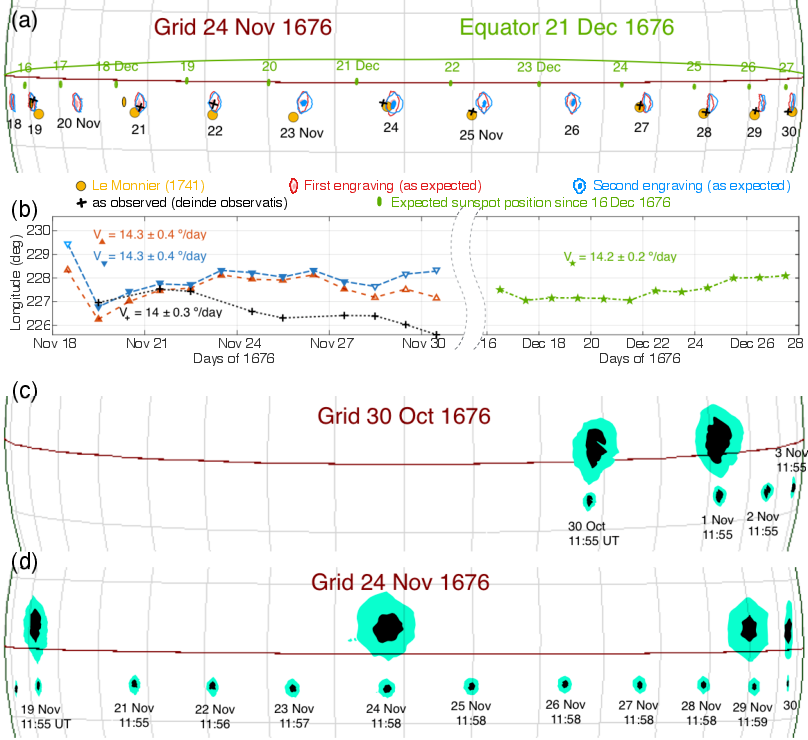}}
\small
        \caption{(\textbf{a}) Superposition of two colored modifications of the original engravings from \citet{Cassini_1676_Nov}. The expected calculated positions of the sunspot in November 1676 are shown in \textit{red} and \textit{blue}. \textit{Green dots} indicate the expected positions in December. \textit{Black crosses} represent the actual observed positions of the spot. Picard's observations are marked with \textit{yellow circles}. (\textbf{b}) Sunspot longitudes and rotation rate shown in various colors. (\textbf{c}) and (\textbf{d}) Our reconstructed visualization of the sunspot’s transit, combining measurements from \cite{Monnier_1741} with sunspot engraving from  \cite{Cassini_1730}.}
\label{Fig12}
\end{figure}

First, we would like to present a loose translation from \citet{Cassini_1730}, who tested his method of predicting the trajectory of the sunspot that appeared on 18 November 1676. He wrote: astronomers have so far depicted the position of sunspots day by day, but never described the line of their motion. This is the third sunspot to appear this year, a year in which they are more frequent than in the 20 previous years. It is the same sunspot that we saw at the end of last month [October], though  we had not observed it earlier due to clouds. Mr. Picard observed it while taking the altitude of the Sun for clock correction on the morning of 30 October. Although, we cannot be certain about the duration of this type of spot (\textit{de cette sorte de Taches}), which often dissipate in a few days when they form again, I believe, however, based on their size --- which is larger than other spots this year --- that they may reappear. To ensure the sunspot could be observed by multiple  astronomers in different locations, I wrote to Mr. Oldenburg (Oldemburg in French) urging readiness for the observations. Having calculated the time of the sunspot's return and sketched the path it should follow based on my method, I searched for it on the morning of the 18th through a 20-foot telescope. I found it so close to the limb and so faint that it could not be seen with the quarter-circle telescopes we use to measure altitudes. That same day, I reported my findings to the Royal Academy, along with a copy of the predicted trajectory for the spot. The path consists of three motions: the rotation of the Sun around its axis, the apparent motion of this axis around the Ecliptic axis, and the variation in the inclination of the Ecliptic relative to the meridian. Consequently, the next track of the sunspot can differ from its previous trajectory. Besides, each part of the spot exhibits its own motion, though this does not notably affect the calculated track. I hasten to publish this notice. There are Scavans, who would appreciate being forewarned of this phenomenon, which we are not able to observe every time we wish, as the Sun does not always have spots that can be observed. We will also provide observations at the Royal Observatory to refine hypotheses about sunspot motions. There is no greater pleasure and precision in observation than when one has hypotheses about what to expect.

This text highlights Cassini's interest in long-lived recurrent sunspots as part of his efforts to develop a theory of sunspot motion, which he first published in \citet{Cassini_1672}. There are several points we would like to note. Firstly, Cassini wrote that in 1676, sunspots appeared more frequently than in the previous 20 years. Secondly, when the active region was near the limb, it appeared smaller size due to projection effect and could not be observed with the telescopes routinely used for measuring the Sun's altitude at the Paris observatory (\textit{``Ie la trouvay si proche du bord \& si mince \`{a} cause de son obliquit\'{e}, qu'il ne fut pas possible de la voir par les lunetes des quarts-de-cercle qui nous servent \`{a} prendre les hauteurs"}). Thus, a 20-foot telescope was required to observe the sunspot on 18 November 1676. This limitation could be due to insufficient contrast for detecting sunspots close to the limb \citep[for details]{1991QJRAS..32...35S, 1993ApJ...411..909S}. However, the crucial difference in the number of sunspot groups in the observations by La Hire and M\"{u}ller \citep{2021SoPh..296..154H, 2021ApJ...909..166H, 2023SoPh..298..113V, 2024IAUS..365..355Z} can not be solely attributed to the lack of contrast near the limb.

Astronomical records such as solar altitude or solar diameter measurements should not be considered equivalent to spotless days \citep{2021MNRAS.506..650H, 2015SoPh..290.2719C, 2016SoPh..291.2493C, 2024MNRAS.528.6280H}. Pendulum adjustments were conducted regularly at the Paris observatory. For example, in June 1676, the acceleration of the pendulum for one day was about $10''$, while in November 1676, it was about $3''$.

Cassini noted that when sunspots form again, they often dissipated within a few days (\textit{``qui se formant de nouveau se dissipent souvent en peu de jours"}) and emphasized that the Sun does not always exhibit observable spots (\textit{``...ne se rencontrant pas toujo\^{u}rs des Taches dans le Soleil, qui se puissent observe"}).

Figure~\ref{Fig12}(a) illustrates the superposition of two colored modifications of the original engravings. The first of them (in red) depicts the expected sunspot track in November 1676 as calculated by Cassini. The second engraving is more detailed, showing the expected sunspot track in November (blue) and in December (marked by green dots). The actual sunspot positions as observed (\textit{deinde observatis}) are represented by black crosses. The edge of this engraving was curved, and we corrected it. For comparison, Picard's measurements are shown as yellow circles. In the western hemisphere, Picard's and Cassini's observations (black crosses and yellow circles) align well in both latitude and longitude. The average latitude derived from the actual sunspot positions is $-4 \pm 0.7^{\circ}$.

Since the exact time of the observations is unknown, we assume they were made at Noon, acknowledging that this assumption introduces higher uncertainty in longitudes. Figure~\ref{Fig12}(b) displays the derived longitudes and rotation rates in color. The expected longitudes (red and blue triangles) diverge toward the solar limbs. The rotation rate derived from the reliable positions (filled triangles) is $14.3 \pm 0.4^{\circ}$ d$^{-1}$. The actual observed longitudes (black crosses) yield a rotation rate of $14 \pm 0.3^{\circ}$ d$^{-1}$ based on all available data. When calculated separately for the eastern and western hemispheres, the rates are $14.2 - 14.3^{\circ}$ d$^{-1}$. Finally, the rotation rate based on the expected sunspot positions in December (green stars) is $14.2 \pm 0.2^{\circ}$ d$^{-1}$. Notably, the actual sunspot longitudes in December 1676 were at least $1.5^{\circ}$ larger (Figure~\ref{Fig9}b and Figure~\ref{Fig10}b).

\citet{Cassini_1676_Nov, Cassini_1730} provided an additional engraving of the sunspot from 30 October to 30 November 1676. By reversing and flipping this engraving and combining it with Picard's measurements, we present our assumption of the sunspot track in Figures~\ref{Fig12}(c) and (d). This reconstruction is used exclusively to estimate the sunspot area.

\section{Discussion}
\label{S-Discussion}

\begin{figure}    
\centerline{\includegraphics[width=0.5\textwidth,clip=]{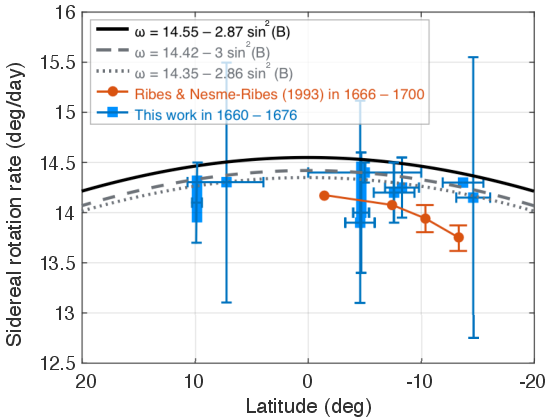}}
\small
        \caption{Sidereal rotation rate from \citet{1993A&A...276..549R} in 1666\,--\,1700, shown in \textit{red}, and those derived in this study, shown in \textit{blue}. The \textit{black solid curve} represents the rotation law derived by \citet{1986A&A...155...87B} for the full range of sunspot groups and the \textit{gray dashed curve} corresponds to the rotation rate for E-F-G-H-J Z\"{u}rich types of sunspot groups. The \textit{gray dotted} curve represents the solar rotation for large long-lived sunspot groups as derived by \citet{2018AstL...44..202N}.}
\label{Fig13}
\end{figure}

Figure~\ref{Fig13} compares the sidereal rotation rates derived in this study. The results from \citet{1993A&A...276..549R} who analyzed observations at the Paris observatory in 1666\,--\,1700, are shown in red. They concluded that the equatorial rate decreased by 2\,--\,3\% during the Maunder minimum \citep{1988IAUS..123..227R}, and that the rotation profile was more differential. The rotation rate obtained from long-living sunspots analyzed in this work are shown in blue. For 1671 and November 1676, the rotation rates are represented by elongated rectangles to account for uncertainties.

We confirm that the derived rotation rates for 1660\,--\,1676 are slightly smaller than the modern rotation profile, represented by the black solid line, which shows the rotation law derived by \citet{1986A&A...155...87B} for the full range of Greenwich sunspot groups. On the other hand, the active regions processed in this study are regular sunspot groups classified as E, F, G, H, and J-types according to the Z\"{u}rich classification. The groups have slower rotation rates, as indicated by the gray dashed curve \citep{1986A&A...155...87B}. A similar rotation profile was also derived by \citet{2018AstL...44..202N} for large long-lived sunspot groups.

In the latitude range of 10\,--\,$20^{\circ}$ south, our study finds higher rotation rates compared to those reported by \citet{1993A&A...276..549R}. However, due to the limited number of data points and substantial uncertainties, we can not conclusively determine whether the rotation profile was more or less differentiated during the Maunder Minimum. We aim to address this question in future research.

\section{Conclusions}
\label{S-Conclusions}

In this study, we analyzed sunspot observations made between 1660 and 1676, focusing on long-lived active regions. We restored the sunspot longitudes and estimated their rotation rates:

\begin{itemize}
\item From Boyle's observations in May\,--\,June 1660, we calculate the rotation rate to be $14.4 \pm 0.1^{\circ}$ d$^{-1}$ at $-5 \pm 5^{\circ}$ of latitude. Extrapolating this estimate on the text reports by Hevelius, covering the period from 22 February to 7 August 1660, we argue that the observed active regions formed an activity nest. The first sunspot group (or groups) persisted for two solar rotations from February to March, while the second magnetic flux portion lasted for three rotations from May to July. The third active region was observed for a single rotation at the beginning of August 1660 and may have belonged to the northern hemisphere. A few other active regions existed for only a few days.

\item From Cassini's measurements and engravings on 11\,--\,19 August 1671, we evaluate the rotation rate at $9.9 \pm 0.5^{\circ}$latitude, which varies between $14.4 \pm 1.5^{\circ}$ d$^{-1}$ and $13.9 \pm 0.6^{\circ}$ d$^{-1}$, depending on sampling.

\item Two schematic engravings by Siverus and one report by Hook from 18 August to 15 September 1671 yield a rotation rate of about $14.3 \pm 0.8^{\circ}$ d$^{-1}$ at $7.3 \pm 3.4^{\circ}$ latitude. The latitude uncertainty prevents us from determining confidently whether this sunspot group was long-lived from 11 August or whether it was part of an activity nest.

\item Based on the measurements by Roemer and Picard from 18 October to 22 November 1672, we calculate a rotation rate of $14.1 \pm 0.7^{\circ}$ d$^{-1}$ at $-14.6 \pm 1.5^{\circ}$ latitude. Matching longitudes indicates that this active region was recurrent, while the text description of its evolution suggests that a new sunspot group emerged in November.

\item In June 1676, the limited measurements by Picard were insufficient to determine the rotation rate.

\item In August 1676, based on Flamsteed's measurements, the rotation rate at $-7.1 \pm 1.4^{\circ}$ latitude is $14 \pm 0.6^{\circ}$ d$^{-1}$ or $14.2 \pm 1^{\circ}$ d$^{-1}$, depending on sampling. Measurements by Halley yield a rotation rate of $14.25 \pm 0.3^{\circ}$ d$^{-1}$ at $-8.3 \pm 1.5^{\circ}$, while those by Picard give $14.2 \pm 0.3^{\circ}$ d$^{-1}$ at $-7.6 \pm 1.8^{\circ}$.

\item From Flamsteed's measurements in October\,--\,November 1676, we evaluate the rotation rate to be $13.9 \pm 0.8^{\circ}$ d$^{-1}$ at $-5.3 \pm 1.4^{\circ}$. From Picard's measurements, we obtain $14 \pm 0.6^{\circ}$ d$^{-1}$ at $-5 \pm 0.7^{\circ}$. From Cassini's engraving, we find the rotation rate to be $14 \pm 0.3^{\circ}$ d$^{-1}$ and 14.2\,--\,$14.3^{\circ}$ d$^{-1}$ at $-4 \pm 0.7^{\circ}$. If this active region, along with that of August 1676, composes an activity nest, its rotation rate may have been about $14.3^{\circ}$ d$^{-1}$.

\end{itemize}

These values are in agreement with the rotation rate of long-lived spots observed in the modern era. However, we would like to emphasize that the derived estimates are subject to several uncertainties, including: (i) duration of observations: ranging from a few minutes to an hour; (ii) solar diameter size: uncertainties reaching up to $10''$; (iii) micrometer and pendulum clock errors introduce a latitude uncertainty of $2-3^{\circ}$ (up to $5^{\circ}$ near the limb) and a longitude uncertainty of $1.5^{\circ}$, both of which  increase toward the limb; (iv) engraving discrepancies yield an additional uncertainty of approximately $2^{\circ}$ in both latitude and longitude; (v) weather conditions such as clouds or wind affected the stability of the observational equipment, contributing to further inaccuracies.

As a minor finding, we note that all astronomers defined a sunspot by its umbra, while referring to penumbra as a cloud. The small black points mentioned by Cassini are likely small, trailing sunspots that were not considered significant enough to be classified as sunspots.

Additionally, due to the projection effect, small sunspots near the limb were reported as unseen with the telescopes used for routine solar altitude measurements. This may explain the discrepancy in the number of sunspot groups reported by the Paris observatory compared to those documented by M\"{u}ller. Cassini, in particular, was interested in long-lived sunspots to develop a theory of sunspot motion. He pointed out that, when sunspots reappeared, they often dissipated within a few days, and that the Sun does not always have observable spots.

For the benefit of open discussion, all processed drawings are available at \url{http://geo.phys.spbu.ru/~ned/History.html}.

\begin{acks}

We use data from the archives of the Royal Society Publishing \url{https://royalsocietypublishing.org/journals}, Bibliotheque Nationale de France \url{https://gallica.bnf.fr}, Biodiversity Heritage Library \url{https://www.biodiversitylibrary.org}, and Bibliotheque l’Observatoire de Paris \url{https://bibnum.obspm.fr}.

\end{acks}

\bibliographystyle{spr-mp-sola}
\bibliography{Zolotova}  

\IfFileExists{\jobname.bbl}{} {\typeout{}
\typeout{****************************************************}
\typeout{****************************************************}
\typeout{** Please run "bibtex \jobname" to obtain} \typeout{**
the bibliography and then re-run LaTeX} \typeout{** twice to fix
the references !}
\typeout{****************************************************}
\typeout{****************************************************}
\typeout{}}

\end{document}